\documentclass[reprint,amsmath,amssymb,amsfonts,aps]{revtex4-1}

\usepackage{graphicx}
\usepackage{dcolumn}
\usepackage{bm}
\usepackage{braket}
\usepackage{color}
\usepackage{amsmath}
\usepackage{mathtools}

\begin{document}

\title{Optimal phase measurements with bright and vacuum-seeded SU(1,1) interferometers}
\author{Brian E. Anderson$^1$, Bonnie L. Schmittberger$^1$,  Prasoon Gupta$^1$, Kevin M. Jones$^2$, and Paul D. Lett$^{1,3}$}
\date{\today}
\affiliation{$^1$Joint Quantum Institute, National Institute of Standards and Technology and the University of Maryland, College Park, MD 20742 USA}
\affiliation{$^2$Department of Physics, Williams College, Williamstown, Massachusetts 01267 USA}
\affiliation{$^3$Quantum Measurement Division, National Institute of Standards and Technology, Gaithersburg, MD 20899 USA}

\begin{abstract}
The SU(1,1) interferometer can be thought of as a Mach-Zehnder interferometer with its linear beamsplitters replaced with parametric nonlinear optical processes. We consider the cases of bright and vacuum-seeded SU(1,1) interferometers using intensity or homodyne detectors. A simplified, truncated scheme with only one nonlinear interaction is introduced, which not only beats conventional intensity detection with a bright seed, but can saturate the phase sensitivity bound set by the quantum Fisher information. 
We also show that the truncated scheme achieves a sub-shot-noise phase sensitivity in the vacuum-seeded case, despite the phase-sensing optical beams having no well-defined phase.
\end{abstract}

\maketitle

\section{Introduction}
Optical phase sensing is a technique used in a wide variety of applications and measurements ranging from materials characterization to medical imaging. The sensitivity of a phase measurement is ultimately limited by the noise. With improving technology, the sensitivity of phase measurement devices, such as interferometers, is no longer limited by technical noise, and is instead limited by the fundamental quantum noise of the interrogating fields. 

One way to improve optical phase shift measurements with an interferometer is to engineer the quantum noise distribution of the fields~\cite{Caves1981}. For example, injection of single-mode squeezed light into the vacuum port of a classical interferometer has been demonstrated to improve the phase sensitivity of the interferometer~\cite{Xiao1987}.  This will be an important aspect of the next generation Laser Interferometer Gravitational-Wave Observatory apparatus~\cite{Aasi2013}.

Interferometers utilizing quantum fields, such as squeezed light, promise an improvement in phase sensitivity over interferometers using coherent states and linear optics. A figure of merit for the performance of an interferometer is its sensitivity relative to the standard quantum limit (SQL). The SQL is commonly defined as the sensitivity $\Delta \phi=1/\sqrt{\bar{N}}$ for a Mach-Zehnder interferometer with a bright coherent state input, where $\bar{N}$ is the average number of photons used in a measurement. One type of interferometer that promises an improvement in phase sensitivity over the SQL is the SU(1,1) class of interferometers.

\begin{figure}
  \centering
  \includegraphics[width=3in]{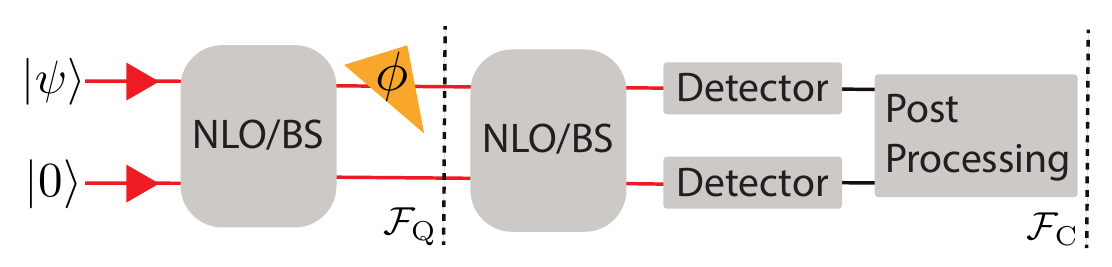}
  \caption
   {A schematic of a Mach-Zehnder or SU(1,1) interferometer. A state $| \psi \rangle$ and a vacuum state $|0\rangle $ are inputs on either a linear beamsplitter (BS) or a nonlinear optical process (NLO). `Unseeded' denotes $| \psi \rangle =|0\rangle $ and `bright seeded' denotes that $| \psi \rangle = | \alpha \rangle$, a coherent state with $|\alpha|^2 \gg 1$. One of the two output beams acquires a phase shift $\phi$ with respect to the other beam and the two beams are recombined in either a BS or NLO. The phase shift $\phi$ is inferred via measurements from the detectors on the final output beams. $\mathcal{F}_\text{Q}$ and $\mathcal{F}_\text{C}$ refer to the Fisher information associated with the quantum state and entire apparatus, respectively (see text).}
\label{fig_simp_schematic}
\end{figure}

The SU(1,1) interferometer can be viewed as a Mach-Zehnder interferometer with its linear beamsplitters replaced with nonlinear optical processes (NLO) that function as parametric gain elements, as shown in Fig~\ref{fig_simp_schematic}. The NLOs can generate photons even when unseeded ($| \psi \rangle = |0\rangle$ in Fig.~\ref{fig_simp_schematic}). It can be shown that an unseeded SU(1,1) type interferometer, under ideal conditions and in the large $\bar{N}$ limit, has a sensitivity $\Delta \phi$ proportional to $1/\bar{N}$ (the Heisenberg limit), thus holding out the promise of a substantial improvement over the SQL. The SU(1,1) interferometer was originally conceived by Yurke \textit{et al.}~in 1986~\cite{Yurke1986}, but only recently has there been progress towards constructing such a device~\cite{Anderson2016,Hudelist2014, Gross2010}. This has raised interest in understanding its operation and phase-sensing ability in more detail. Here we consider the phase-sensing ability of various modifications to the original proposal and show that under some conditions one can simplify the original scheme without reducing the phase sensitivity of the device.

\begin{figure*}
  \centering
  \includegraphics[width=5.4in]{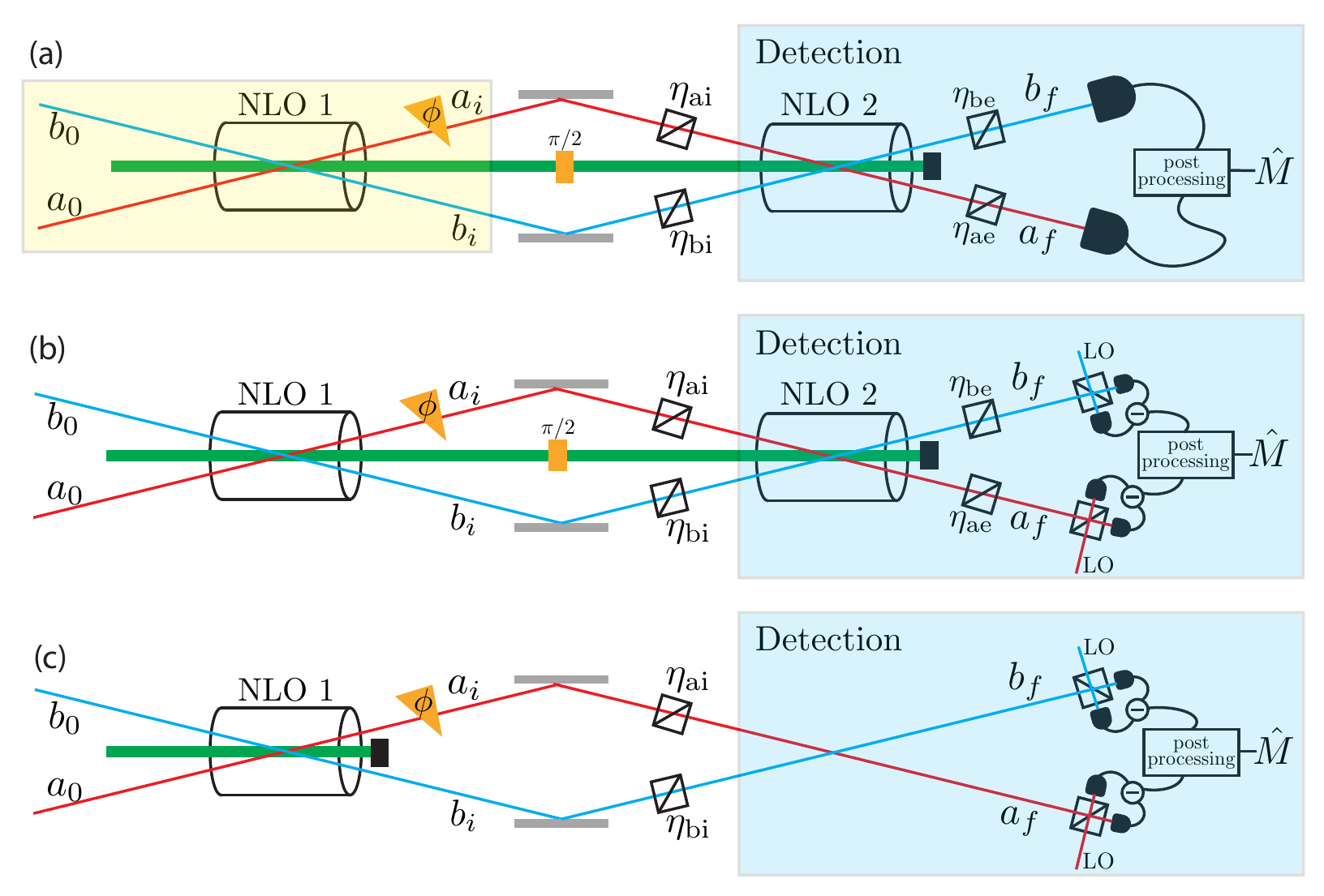}
  \caption
   {The three experimental configurations considered in this paper. The thick green line is a strong classical pump beam. (a) The conventional SU(1,1) interferometer. (b) The conventional SU(1,1) with homodyne detection on the output. (c) The truncated SU(1,1) interferometer. LO indicates a local oscillator, and beamsplitters represent loss and are labelled by their transmission $\eta$.}
\label{fig_schematics}
\end{figure*}

To analyze the phase sensitivity of an interferometer it is useful to break the operation into two stages: The first is the production of a quantum state, a portion of which passes through the phase object. The second stage is the measurement of some aspect of this quantum state, from which the phase shift is inferred. An important development in our understanding of quantum sensing is the realization that analysis of the first stage alone, independent of the second stage,  sets limits on the potential phase sensitivity of the interferometer~\cite{Sparaciari2016}. These limits on the best phase sensitivity of an interferometer can be quantified using the Fisher information. 

The Fisher information is a well-established metric for measuring how much information a statistical random variable carries about an unknown parameter~\cite{Trees2001}. We calculate a Fisher information associated with the quantum state, $\mathcal{F}_\text{Q}$, immediately after the phase object, as shown in Fig.~\ref{fig_simp_schematic}. We can also calculate a Fisher information associated with the entire apparatus including detection, $\mathcal{F}_\text{C}$, which is necessarily less than or equal to $\mathcal{F}_\text{Q}$. These are conventionally referred to as the quantum ($\mathcal{F}_\text{Q}$) and classical ($\mathcal{F}_\text{C}$) Fisher information~\cite{Sparaciari2016}. 

The goal of this work is to find the best detection apparatus and signal processing such that $\mathcal{F}_\text{C}$ approaches the limit set by $\mathcal{F}_\text{Q}$, and thus the interferometer achieves a sensitivity limited only by the quantum state used to sense the phase object. The two-mode squeezed state produced by the first NLO gives SU(1,1) interferometers a potential improvement over the SQL, but the choice of detection apparatus is crucial to realize that improvement. We will discuss several measurement configurations, as shown in Fig.~\ref{fig_schematics}: conventional SU(1,1) with optical intensity measurement (Fig.~\ref{fig_schematics}a); conventional SU(1,1) with optical homodyne measurement (Fig.~\ref{fig_schematics}b); and ``truncated" SU(1,1) with optical homodyne measurement (Fig.~\ref{fig_schematics}c). We focus on optical intensity and homodyne measurements because these are the primary detection tools available to experimentalists. In some potential physical realizations of the SU(1,1) interferometer, scattered light from the intense pump beam can make direct intensity detection of weak signal beams problematic.  Homodyne detection, which provides strong spectral filtering, can help overcome this problem.
 
We organize the paper as follows. In Sec.~\ref{sec_sensi_FI}, we provide a background on sensitivity measurements and Fisher information~\cite{Sparaciari2015}. In Sec.~\ref{sec_su11_int}, we discuss the sensitivity for the conventional SU(1,1) system using optical intensity detection, which has been discussed elsewhere~\cite{Plick2010a,Chen2016,Pezze2015,Ou2012,Gong2016,Marino2012a,Hudelist2014,Li2014,Kong2013}. We show that intensity detection is the optimal detection scheme for the vacuum-seeded case, but is not optimal in the bright-seeded case. In Sec.~\ref{sec_su11_hd}, we discuss the sensitivity for SU(1,1)-type interferometers that use optical homodyne detection, and describe a simplified (truncated) scheme that allows one to measure sensitivities that saturate the limit set by the Fisher information in the bright-seeded case. In Sec.~\ref{sec_sensi_seed}, we discuss the sensitivity, using different detection schemes, as a function of seed power. Finally, in Sec.~\ref{sec_ext_loss_gain}, we discuss modifying the nonlinear gain of the second NLO to improve phase sensitivity in the case that optical detection efficiency is not ideal.

\section{Sensitivity and Fisher Information}
\label{sec_sensi_FI}
We define the sensitivity, $\Delta \phi$, as the uncertainty in estimating a phase shift $\phi$ according to
\begin{equation}
\Delta^2 \phi= \frac{\Delta^2 \hat{M}}{\left(\partial_{\phi} \braket{\hat{M}}\right)^2},
\label{eq_sensi}
\end{equation}
given a measurement represented by an observable $\hat{M}$. The variance of the measurement is $\Delta^2 \hat{M} = \braket{\hat{M}^2}-{\braket{\hat{M}}}^2$. For example, a common choice of observable is the number of photons, $\hat{M} = \hat{a}^{\dagger} \hat{a}$, in an optical beam. There is a Fisher information associated with any measurement, which quantifies how much information about an unknown optical phase shift is contained in the measurement results. Every Fisher information implies a bound, called the Cramer-Rao bound, on the minimum phase shift, $\Delta \phi$, that can be resolved using that information. The Cramer-Rao bound is equal to the inverse of the square root of the Fisher information, which implies that $\Delta \phi \geq 1/\sqrt{\mathcal{F}}$. It is conventional to call the Fisher information associated with a measurement the classical Fisher information, $\mathcal{F}_\text{C}$, and its associated bound the classical Cramer-Rao bound (CCRB). This is not a fundamental limit, however. $\mathcal{F}_\text{C}$ is bounded by the quantum Fisher information, $\mathcal{F}_\text{Q}$, which describes the information associated with parameter estimation limited only by the quantum state. The bound associated with $\mathcal{F}_\text{Q}$ is the quantum Cramer-Rao bound (QCRB). The $\mathcal{F}_\text{Q}$ can be viewed as  $\mathcal{F}_\text{C}$ optimized over measurement schemes, such that $\Delta \phi \ge 1/\sqrt{\mathcal{F}_\text{C}} \ge 1/\sqrt{\mathcal{F}_\text{Q}}$. $\mathcal{F}_\text{Q}$, however, is independent of the measurement scheme and can be calculated using the quantum state alone.

For Gaussian states (those with Gaussian distributions in phase space) and Gaussian measurements (those with a Gaussian distribution for their outcomes), $\mathcal{F}_\text{C}$ can be written as
\begin{equation}
\mathcal{F}_\text{C} = \frac{\left(\partial_{\phi} \braket{\hat{M}}\right)^2}{\Delta^2 \hat{M}} + \frac{2 \left[\partial_{\phi} (\Delta \hat{M})\right]^2}{\Delta^2 \hat{M}},
\label{eq_CFI}
\end{equation}
as shown in Ref.~\cite{Sparaciari2016}. The first term of Eq.~(\ref{eq_CFI}) describes the change in mean value of the measurement, and the second term describes the change in the standard deviation of the distribution. The first term is also identical to the inverse of $\Delta^2 \phi$. In this work, we only consider Gaussian states. Therefore in the case of Gaussian measurements, like homodyne detection, Eq.~(\ref{eq_CFI}) applies. Measurements of photon number are not Gaussian (they are bounded below by zero), so Eq.~(\ref{eq_CFI}) is not applicable.

For Gaussian states, $\mathcal{F}_\text{Q}$ can be calculated with a straightforward procedure if the covariance matrix of the quantum state is known~\cite{Sparaciari2016,Chen2016}. One could include the effect of losses on the quantum state as part of $\mathcal{F}_{\text{Q}}$, but we will maintain $\mathcal{F}_{\text{Q}}$ as an idealized limit in a system with no imperfections. Consider just the portion of the apparatus shown in the yellow box around NLO 1 of Fig.~\ref{fig_schematics}a. Two modes,  $a_0$ and $b_0$, are coupled in an NLO with an intense classical pump beam. After the NLO, mode $a_i$ acquires a phase shift with respect to mode $b_i$. We now consider the bright-seeded case where mode $a_0$ is a coherent state with mean photon number  $|\alpha|^2 \gg 1$ and mode $b_0$ is vacuum. In the ideal lossless case, the quantum Fisher information associated with the two-mode state $a_i$ and $b_i$ is \cite{Sparaciari2016}
\begin{equation}
\mathcal{F}^{\text{NLO}}_{\text{Q}} = 2 \cosh ^2(r) \left[(2 \left| \alpha \right| ^2  +1) \cosh (2 r)-1\right], 
\label{eq_QFI}
\end{equation}
where $r$ is the squeezing parameter, which is related to the intensity gain, $g$, of the NLO process by $g=\cosh^2(r)$. In the limit that $|\alpha|^2 \gg 1$, mode $a_i$ will have mean photon number $g |\alpha|^2$ and mode $b_i$ will have mean photon number $g |\alpha|^2 - |\alpha|^2$. When $g> 1$, modes $a_i$ and $b_i$ are quantum correlated and constitute a two-mode squeezed state. Equation~(\ref{eq_QFI}), via the relation $\Delta \phi \ge 1/\sqrt{\mathcal{F}_\text{Q}}$, determines the maximum possible sensitivity of the interferometer independent of the detection scheme used to infer the phase shift. 

It may seem that the presence of the second NLO in Fig.~\ref{fig_schematics}a may alter the fundamental phase sensitivity of the entire interferometer, but this is not true. The second NLO is a unitary process, thus $\mathcal{F}_{\text{Q}}$ calculated for modes $a_i$ and $b_i$ is the same as for modes $a_f$ and $b_f$. Conceptually, the second NLO neither adds information about the phase shift that came before it, nor can it subtract information since it is a unitary process. This can be seen in a different way by considering $\mathcal{F}_\text{Q}$ as an optimization of $\mathcal{F}_\text{C}$ over all positive-operator valued measurements (POVMs). All POVMs can be reduced to a unitary process in a Hilbert space larger than the original quantum state, followed by a von Neumann measurement~\cite{Nielsen2000}. Since the second NLO is a unitary transformation itself, we can include it as part of the measurement apparatus, and we show this in Fig.~\ref{fig_schematics} by including the second NLO as part of the detection stage. This observation allows us to remove the second nonlinear interaction, as shown in Fig.~\ref{fig_schematics}c, while preserving the inherent phase sensitivity of the device. This truncated SU(1,1) interferometer will be discussed further in Sec.~\ref{sec_su11_hd}.

\subsection{The Standard Quantum Limit}
Ultimately, the goal of SU(1,1) and other quantum-enhanced interferometers is to measure phase with a sensitivity better than the SQL. As shown in Fig. 2, SU(1,1) interferometers have several classical phase references not found in a conventional Mach-Zehnder interferometer: pump beams driving the NLOs and possibly local oscillators (LOs) for optical homodyne detection.  In addition, for a bright-seeded SU(1,1) interferometer, modes $a_i$ and $b_i$ have different mean photon numbers, unlike a conventional balanced Mach-Zehnder interferometer where the two arms have equal photon numbers.  These features open the possibility of different choices of linear-optical interferometers operating with coherent beams, with which to define the SQL.

\begin{figure}
  \centering
  \includegraphics[width=3in]{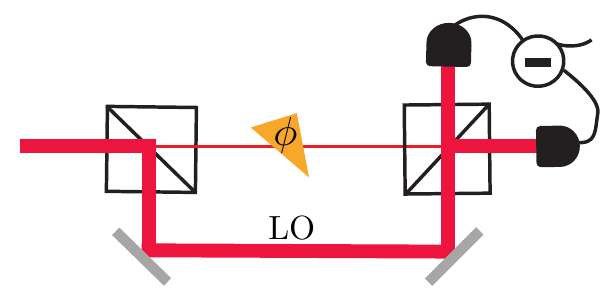}
  \caption
   {A phase sensor that can be viewed either as an imbalanced Mach-Zehnder interferometer or as a balanced homodyne detector with a local oscillator (LO) that has a well-defined phase with respect to the weak beam. The weak beam passing through the optical phase shift, $\phi$, contains $\bar{N}_p$ photons on average.}
\label{fig_schematics_ph}
\end{figure}

A standard definition of the SQL is the sensitivity of a Mach-Zehnder, $\Delta \phi=1/\sqrt{2\bar{N}_p}$, with mean number $2\bar{N}_p$ photons seeding the interferometer and $\bar{N}_p$ photons in one arm passing through a phase object. However, consider the sensitivity of the device in Fig.~\ref{fig_schematics_ph}, that can be viewed either as a balanced homodyne detector or an unbalanced Mach-Zehnder interferometer. If $\bar{N}_p$ photons pass through the phase object, the sensitivity of this device is $\Delta \phi=1/\sqrt{4 \bar{N}_p}$. The discrepancy between this sensitivity and the standard SQL above comes from the device in Fig.~\ref{fig_schematics_ph} having a local oscillator (LO) with an optical power much larger than the weak beam that acts as an external phase reference~\cite{Jarzyna2012}. The shot noise of the nearly classical LO does not affect the phase sensitivity, so one achieves a factor of $\sqrt{2}$ improvement in sensitivity compared to a balanced Mach-Zehnder. This ``enhancement" over the standard definition of SQL is a matter of definition, rather than a quantum effect.

We will adopt the definition that only the number of photons passing through the phase object will be counted when defining the SQL. This is consistent with the idea that one is, for example, trying to measure the phase shift of a sample which is subject to optical damage. Given that, we take the most conservative classical bound possible: the sensitivity of the device in Fig.~\ref{fig_schematics_ph}, $(\Delta \phi)_{\text{SQL}}=1/\sqrt{4 \bar{N}_p}$, with $\bar{N}_p$ photons passing through the sample. Other authors have defined the SQL differently~\cite{Hudelist2014}.

Returning now to Eq.~(\ref{eq_QFI}), it's instructive to consider the $r=0$ limit for which $g=1$ and there is no quantum enhancement. In this case, mode $a_i$ is a coherent state with mean photon number $| \alpha|^2$, mode $b_i$ is vacuum and $\mathcal{F}_{\text{Q}} = 4 | \alpha|^2$. This is simply the Fisher information associated with a single coherent state modified by a phase shift, $\phi$, as shown in Fig.~\ref{fig_schematics_ph} and discussed above. We can re-express  the SQL defined above in terms of a Fisher information, which we designate as $\mathcal{F}^{\text{SQL}}_{\text{Q}} = 4 \bar{N}_p$.

We can now compare $\mathcal{F}^{\text{NLO}}_{\text{Q}}$ and $\mathcal{F}^{\text{SQL}}_{\text{Q}}$ for the SU(1,1) interferometer as a function of gain. Only mode $a_i$ is modified by $\phi$ and has $\bar{N}_p = |\alpha|^2 \cosh^2(r) + \sinh^2(r)$ photons on average, so
\begin{equation}
\mathcal{F}^{\text{SQL,r}}_{\text{Q}} = 4[ |\alpha|^2 \cosh^2(r) + \sinh^2(r)].
\label{eq_QFI_MZ}
\end{equation}
The SQL depends not only on the seed intensity but also on the gain of the NLO, because increasing gain  increases the intensity of light in the interferometer. Assuming, as before, a bright seed ($|\alpha|^2 \gg 1$), the ratio $\mathcal{F}^{\text{NLO}}_{\text{Q}} / \mathcal{F}^{\text{SQL,r}}_{\text{Q}}= \cosh(2r)$. The benefit of the SU(1,1) interferometer improves with increasing $r$, and at $g=1$ ($r=0$), the SU(1,1) interferometer operates at the SQL as defined above. With this background, we now turn to an analysis  of the different detection schemes shown in Fig.~\ref{fig_schematics} and compare their sensitivities to the QCRB set by the $\mathcal{F}_{\text{Q}}$ given in Eq.~\ref{eq_QFI}.

\section{SU(1,1) Interferometer with Intensity Detection}
\label{sec_su11_int}
To begin our analysis, we consider the experiment shown in Fig. \ref{fig_schematics}a, which shows an SU(1,1) interferometer with intensity detection. This configuration has been studied previously~\cite{Hudelist2014,Ou2012,Gong2016,Marino2012a}, so we will just provide basic results for use in comparing to other configurations.

The first NLO generates two-mode squeezing with the two input modes, $a_0$ and $b_0$, coupled by an intense classical pump beam. Mode $a_i$ acquires a phase $\phi$ with respect to the other mode and the two modes are then coupled in NLO 2 with an intense classical pump. We will assume, until Sec.~\ref{sec_ext_loss_gain}, that both NLOs are characterized by the same magnitude squeezing parameter. The second NLO is a phase-sensitive device whose output depends on the relative phases of the input beams, including the classical pump, $2\phi_{\text{pump}} - \phi_{\text{a}}  - \phi_{\text{b}}$, where $\phi_{\text{a}}$ is the phase of mode $a_i$, $\phi_{\text{b}}$ is the phase of mode $b_i$, and $\phi_{\text{pump}}$ is the phase of the pump. We choose that the pump phase is shifted by $\pi/2$ after the first NLO such that if $\phi=0$, the unitary transformation performed by the second NLO is the inverse of the first NLO. We represent loss and detector inefficiency with beamsplitters between the NLOs and after the second NLO, respectively.

\subsection{Bright Seed}

\begin{figure}
  \centering
 \includegraphics[height=5.5cm]{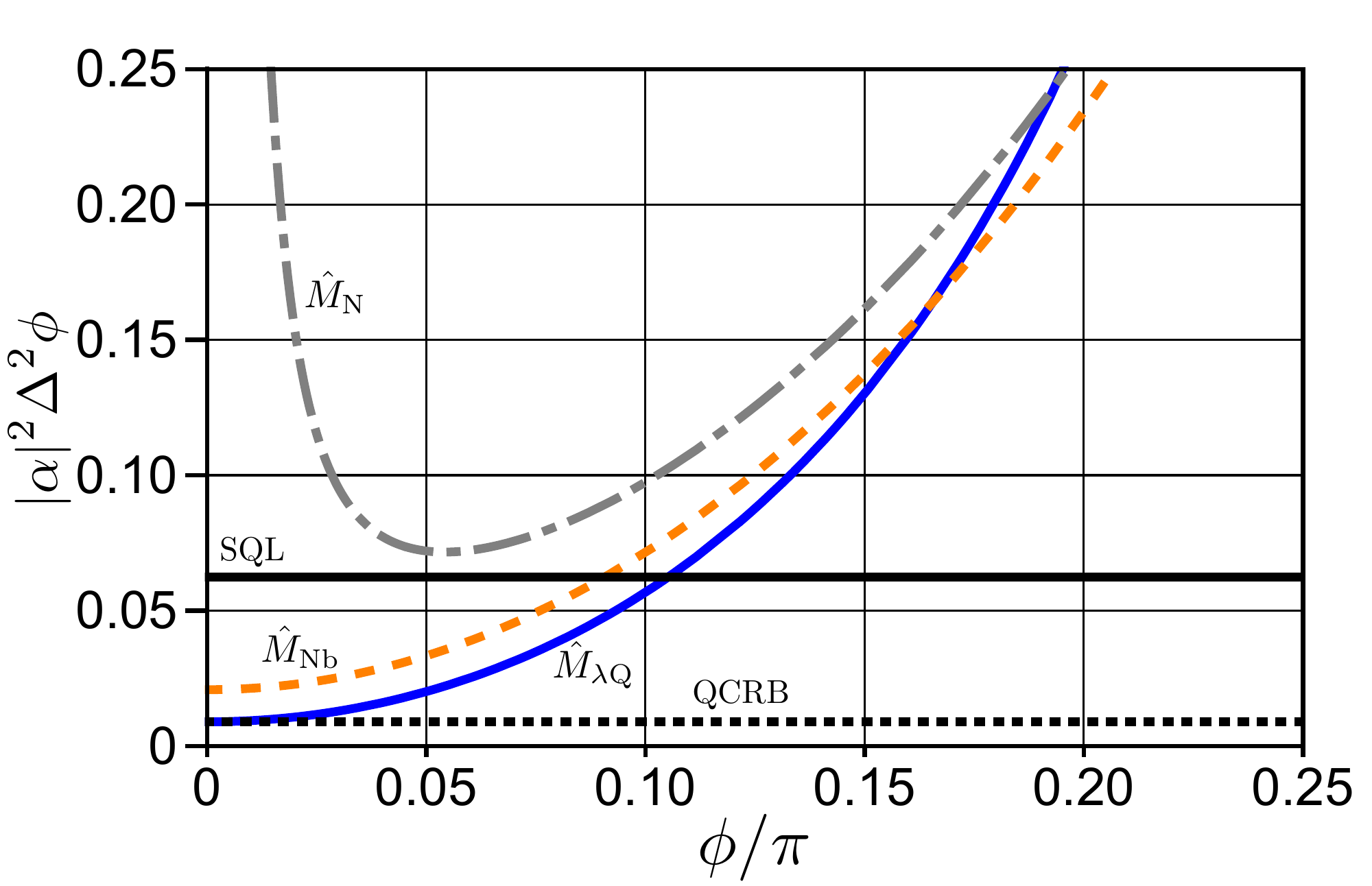}\llap{\raisebox{1cm}{\includegraphics[height=2cm]{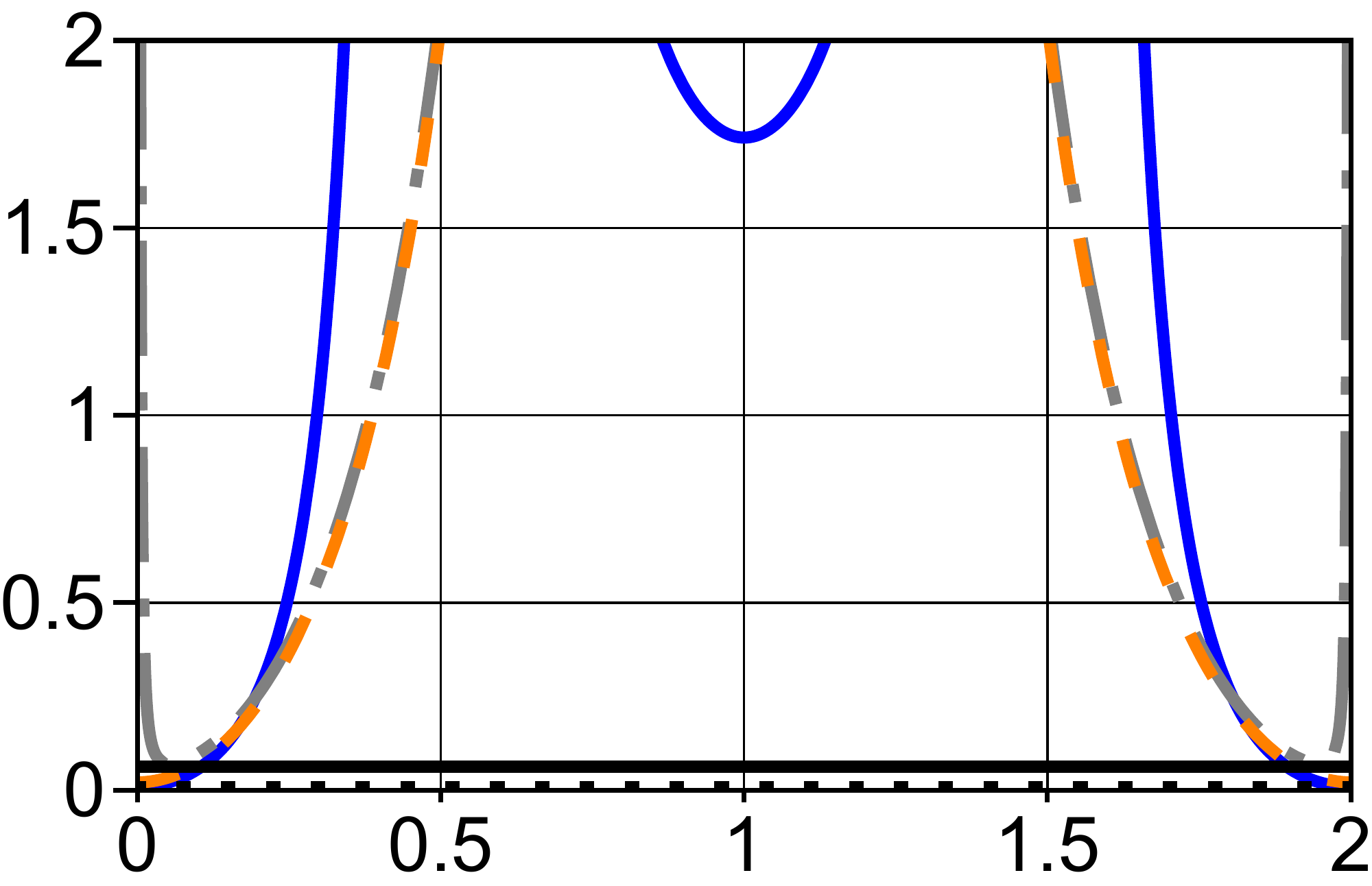}}}
  \caption
   {Sensitivity as a function of operating point for bright-seeded configurations with a gain of 4. Sensitivity is shown for several experimental configurations: intensity detection ($\hat{M}_{\text{N}}$) is gray dash-dot, intensity detection of mode $b_f$ only ($\hat{M}_{\text{Nb}}$) is orange dashed, homodyne detection for truncated and conventional SU(1,1) interferometer  ($\hat{M}_{\text{Q}}$) with LO phases fixed at $\theta_a =\theta_b =  \pi/2$ is thick blue, and homodyne detection for truncated SU(1,1) interferometer with classical gain correction ($\hat{M}_{\lambda \text{Q}}$) optimized over LO phase is black dotted. The QCRB coincides with black dotted and the SQL is solid black.  The inset shows the same graph over a larger range.}
\label{fig_bright_vs_theta}
\end{figure}

\begin{figure}
  \centering
  \includegraphics[width=3.3in]{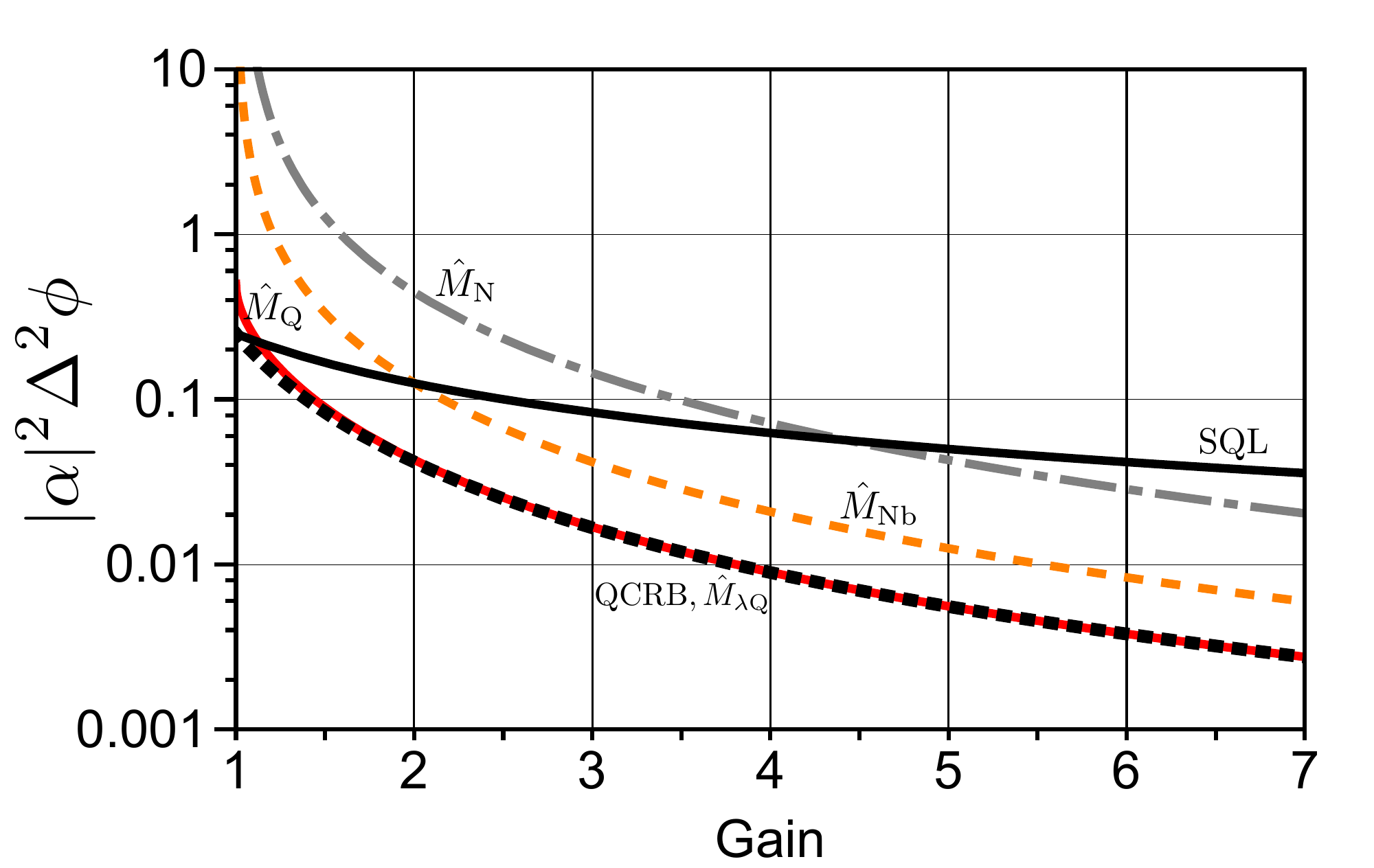}
  \caption
   { Sensitivity as a function of gain for bright seeded configurations, optimized over $\phi$ and LO phases. Sensitivity is shown for several experimental configurations: intensity detection ($\hat{M}_{\text{N}}$) is gray dash-dot, intensity detection of mode $b_f$ only ($\hat{M}_{\text{Nb}}$) is orange dashed, homodyne detection for truncated and conventional SU(1,1) interferometer ($\hat{M}_{\text{Q}}$) is solid red, and homodyne detection for truncated SU(1,1) interferometer with classical gain correction ($\hat{M}_{\lambda \text{Q}}$) is black dotted. The QCRB coincides with black dotted and the SQL is solid black.}
\label{fig_bright_comp}
\end{figure}

In the bright-seed configuration, mode $a_0$ is a coherent state with mean photon number $|\alpha|^2 \gg 1$ and mode $b_0$ is the vacuum. If we take as the signal the sum of the two output intensities, then $\hat{M}_{\text{N}} = \hat{a}^{\dagger}_f \hat{a}_f + \hat{b}^{\dagger}_f \hat{b}_f$, and we find
\begin{equation}
(\Delta^2 \phi)^{\text{S}}_{\text{N}}  = \frac{\text{csch}^4(2 r) \left[\cosh (8 r) \sec ^2\left(\frac{\phi }{2}\right)+\csc ^2\left(\frac{\phi }{2}\right)\right]-8}{4 |\alpha|^2 }
\label{eq_num_sensi}
\end{equation}
in the case of no loss. In this notation, the subscript of $\Delta^2 \phi$ denotes measurement choice and the superscript denotes bright-seeded (S) or unseeded (U). To perform this calculation, we use the formalism found in \cite{Anderson2016, Plick2010a}, which allows us to calculate $\hat{a}_f$ and $\hat{b}_f$ in the Heisenberg picture. Analytic expressions for these operators allowed us to calculate expectation values for various measurements, as well as the covariance matrix. We plot Eq.~(\ref{eq_num_sensi}) as a function of $\phi$ in Fig.~\ref{fig_bright_vs_theta}. In all bright-seeded cases, $\Delta^2 \phi$ is proportional to $1/|\alpha|^2$, so we rescale the vertical axis by $|\alpha|^2$.  Only the gray dash-dot line is relevant here, and other lines are discussed in later sections. The minimum of Eq.~(\ref{eq_num_sensi}) occurs at a phase shift
\begin{equation}
\phi = 2 \cot ^{-1}\left[\sqrt[4]{\cosh (8 r)}\right],
\end{equation}
which gives an optimum $\Delta^2 \phi$ of
\begin{equation}
(\Delta^2 \phi)^{\text{S}}_{\text{N}} \Big |_{\text{min}} =\frac{\left[2 \cosh (4 r)+\sqrt{\cosh (8 r)}-1\right]\text{csch}^4(2 r)}{2 |\alpha|^2 }.
\label{eq_num_opt}
\end{equation}
The expression in Eq.~(\ref{eq_num_opt}) never saturates the QCRB, which is shown in Fig.~\ref{fig_bright_vs_theta} as a black dotted line. Using the relationship between $r$ and gain, we plot the optimal $\Delta^2 \phi$ as a function of gain in Fig.~\ref{fig_bright_comp}. 

We can write the squeezing parameter, $r$, in terms of $\bar{n}_s$, the mean number of spontaneous photons in both modes $a_i$ and $b_i$ when there is no seed, which is $\bar{n}_s = 2\sinh^2(r) = 2g-2$. Re-writing Eq.~(\ref{eq_num_opt}) in terms of these photons, we have
\begin{multline}
(\Delta^2 \phi)^{\text{S}}_{\text{N}}\Big |_{\text{min}} = \\
 \frac{4 \bar{n}_s (\bar{n}_s+2)+\sqrt{\cosh \left[8 \sinh
   ^{-1}\left(\sqrt{\frac{\bar{n}_s}{2}}\right)\right]}+1}{2 |\alpha|^2  \bar{n}_s^2 \left(\bar{n}_s+2\right)^2},
\end{multline}
and the scaling of this quantity is $1 / |\alpha|^2 \bar{n}^2_s$ in the large $\bar{n}_s$ limit. For the SQL, the sensitivity $\Delta \phi$ scales as $1/ \sqrt{\bar{N}}$, but for the SU(1,1) there are two different scalings: In terms of seed photon number, the sensitivity scales as $1/ \sqrt{\bar{N}}$, whereas in terms of spontaneous photon number, the sensitivity scales as $1/\bar{N}$. The improved scaling property of SU(1,1) interferometers has been lauded as a motivation for their study, but the scaling improvement only appears in the spontaneous photon number, rather than all photons. The coherent seed is simply a multiplicative factor in the sensitivity~\cite{Plick2010a}. Furthermore, given even modest loss in the apparatus, the scaling quickly approaches $1/\sqrt{\bar{N}}$ even for spontaneous photons~\cite{Marino2012a}.

An alternative detection scheme to measuring the sum of the two output intensities is to measure the intensity only of output $b_f$ so that $\hat{M}_{\text{Nb}}= \hat{b}^{\dagger}_f \hat{b}_f $, giving a sensitivity of
\begin{equation}
(\Delta^2 \phi)^{\text{S}}_{\text{Nb}} =\frac{\cosh (4 r) \text{csch}^2(r) \text{sech}^2(r) \sec ^2\left(\frac{\phi }{2}\right)-8}{4 |\alpha|^2 }.
\label{eq_M_nb}
\end{equation}
This produces considerably different behavior compared to Eq.~\ref{eq_num_sensi} as a function of operating point, as shown in Fig.~\ref{fig_bright_vs_theta}. When $\phi \rightarrow 0$, $b_f$ becomes the vacuum, and the numerator and denominator of Eq.~(\ref{eq_sensi}) go to zero at the same rate, giving the well-defined limit
\begin{equation}
(\Delta^2 \phi)^{\text{S}}_{\text{Nb}}\Big |_{\text{min}} = \frac{\text{csch}^2(2 r)}{|\alpha|^2 }.
\label{eq_M_nb_min}
\end{equation}
The behavior changes qualitatively in the presence of loss, and the sensitivity diverges as $\phi \rightarrow 0$, causing the orange dashed curve to turn up similar to the gray dash-dot. One can see in Figs.~\ref{fig_bright_vs_theta} and \ref{fig_bright_comp} that neither $\hat{M}_{\text{N}}$ nor $\hat{M}_{\text{Nb}}$ saturates the QCRB. We show in Sec.~\ref{sec_su11_hd} how to saturate this bound in the bright-seeded case.

\subsection{Vacuum Seed}

\begin{figure}
  \centering
  \includegraphics[width=3.3in]{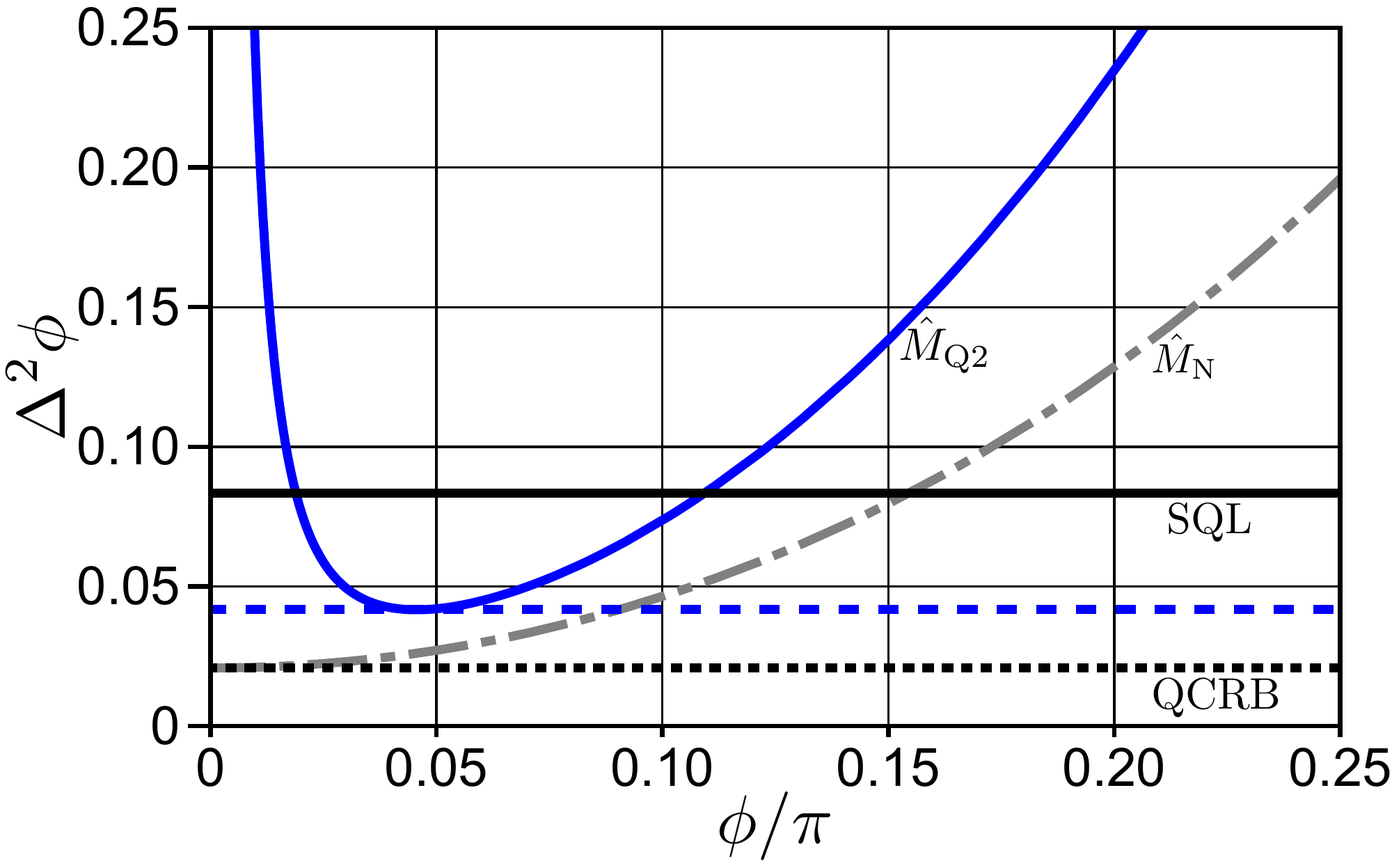}
  \caption
   { Sensitivity as a function of operating point for vacuum-seeded schemes with a gain of 4. Sensitivity is shown for several experimental configurations:  intensity detection ($\hat{M}_{\text{N}}$) is in gray dash-dot, homodyne detection of truncated and conventional SU(1,1) interferometer ($\hat{M}_{\text{Q2}}$) with LOs set at $\theta_a = \theta_b = \pi/2$ is dark blue, and homodyne detection of truncated and conventional SU(1,1) ($\hat{M}_{\text{Q2}}$) optimized over LO phases is blue dashed. The QCRB is the black dotted line and the SQL is the solid black line.}
\label{fig_vacuum_vs_theta}
\end{figure}

A different variation on the SU(1,1) with intensity detection is to let the first NLO be vacuum-seeded. If both modes $a_0$ and $b_0$ are vacuum we find that for the measurement $\hat{M}_{\text{N}}$,
\begin{equation}
(\Delta^2 \phi)^{\text{U}}_{\text{N}} = \coth ^2(2 r) \sec ^2\left(\frac{\phi }{2}\right)-1,
\end{equation}
which has a minimum when $\phi \to 0$. At this phase, the optimal sensitivity is
\begin{equation}
(\Delta^2 \phi)^{\text{U}}_{\text{N}}\Big |_{\text{min}} = \text{csch}^2(2 r).
\end{equation}
In this vacuum-seeded case, $(\Delta^2 \phi)^{\text{U}}_{\text{N}}|_{\text{min}} $ saturates the QCRB, as shown in Fig.~\ref{fig_vacuum_vs_theta}, and $\hat{M}_{\text{N}}$ is an optimal measurement. Similar to the measurement $\hat{M}_{\text{Nb}}$ with a bright seed in the previous section, any loss qualitatively changes the behavior of the sensitivity curve and causes the gray dash-dot line in Fig.~\ref{fig_vacuum_vs_theta} to turn up as $\phi \rightarrow 0$. We saturate the QCRB in the seeded case by using a homodyne detection scheme discussed below.

\section{SU(1,1) Interferometer with Homodyne Detection}
\label{sec_su11_hd}
In this section, we describe detection schemes using optical homodyne detectors to measure selected quadratures of the interferometer's output modes, as shown in Figs.~\ref{fig_schematics}b and \ref{fig_schematics}c. The scheme in Fig.~\ref{fig_schematics}b is the conventional SU(1,1) interferometer with homodyne detection~\cite{Manceau2017,Li2014}. We begin, however, by considering the experiment shown in Fig. \ref{fig_schematics}c, which we call the ``truncated" SU(1,1) interferometer. Although the setup looks like an incomplete interferometer, homodyne detection is a phase sensitive measurement by itself. The device therefore has an associated phase sensitivity given that the homodyne local oscillator (LO) phases (that determine the phase of the quadrature measurement) are locked relative to the phase of the bright seed. In the case of a vacuum seed, the LOs could be locked to the pump.

\begin{figure}
  \centering
  \includegraphics[width=3.3in]{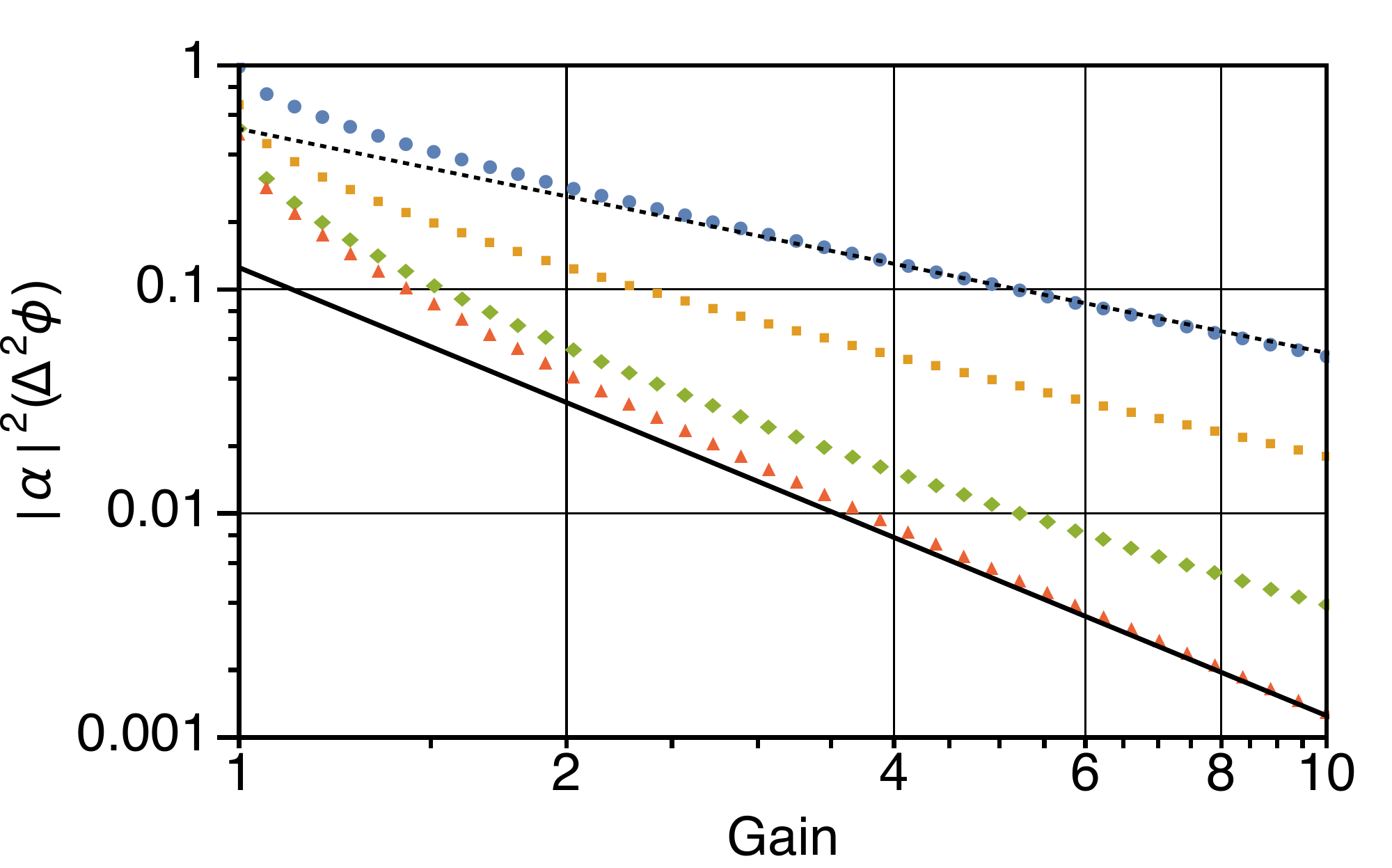}
  \caption
  {Sensitivity as a function of gain for homodyne measurement of bright seeded truncated or conventional SU(1,1) interferometer ($\hat{M}_{\text{Q}}$) and no external loss. Gain is linear in number of spontaneous photons: $\bar{n}_s = 2g - 2$. The blue dots, yellow squares, green diamonds, and orange triangles represent internal transmissions $\eta_{\text{ai}} = \eta_{\text{bi}} = 0.5, 0.75, 0.95$ and $1$, respectively. The black lines are guides for the eye where the dashed has scaling $1/g$ and the solid has scaling $1/g^2$.}
\label{fig_bright_hd_scaling}
\end{figure}

\subsection{Bright Seed}
We take the output of the two homodyne detectors and add them, which gives 
\begin{equation}
\hat{M}_{\text{Q}} = e^{ i \theta_a}\hat{a}_f + e^{ -i \theta_a}\hat{a}^{\dagger}_f + e^{ i \theta_b}\hat{b}_f + e^{ -i \theta_b}\hat{b}^{\dagger}_f,
\end{equation}
where $\theta_a$ and $\theta_b$ are the LO phases for modes $a_f$ and $b_f$, respectively. We take $\alpha$ to be real so that an LO phase of $\theta_a=\pi/2$ or $\theta_b=\pi/2$ corresponds to measuring the phase quadrature. Taking $\theta_a = \pi/2$ in the case of no loss and $|\alpha|^2 \gg 1$, we calculate 
\begin{multline}
(\Delta^2 \phi)^{\text{S}}_{\text{Q}}  = \\
\frac{\sec ^2(\phi ) \left[1-2 \tanh (r) \sin (\theta_b-\phi )+\tanh ^2(r)\right]}{2 \left| \alpha \right| ^2 },
\label{eq_trunc_hd_sensi_ph}
\end{multline}
and plot in Fig.~\ref{fig_bright_vs_theta}. In the presence of loss, unlike the intensity detection curves discussed earlier, the sensitivity curve does not diverge as $\phi \rightarrow 0$. If we optimize over LO phase $\theta_b$ and $\phi$, the optimum sensitivity value is
\begin{equation}
(\Delta^2 \phi)^{\text{S}}_{\text{Q}}\Big |_{\text{min}}  =\frac{\left[\tanh (r)-1\right]^2}{2 \left| \alpha \right| ^2}.
\label{eq_trunc_hd_sensi}
\end{equation}
This operating point is achieved when $\theta_b=\pi/2$, i.e.~both homodyne detectors are set to detect the phase quadrature of their respective input beams, and the measurement $\hat{M}_{\text{Q}}$ is the phase quadrature sum. We plot Eq.~(\ref{eq_trunc_hd_sensi}) in Fig.~\ref{fig_bright_comp}.

Analysis of the sensitivity of the conventional SU(1,1) interferometer with homodyne detection, as shown in Fig.~\ref{fig_schematics}b, shows that it achieves the same phase sensitivity as the truncated SU(1,1) interferometer in Eq.~(\ref{eq_trunc_hd_sensi}). The optimal operating point of the conventional interferometer, with respect to the LO phases and the internal phase shift, is somewhat more complicated, but the optimal sensitivity is identical. 

For conventional and truncated SU(1,1) interferometer systems using homodyne detection, in the limit that $|\alpha|^2 \gg 1$, the second term of Eq.~(\ref{eq_CFI}) is negligible everywhere and zero at the optimal point, so the sensitivity in Eq.~(\ref{eq_trunc_hd_sensi}) saturates the CCRB. For low gain, the CCRB does not quite reach the QCRB (the red curve lies slightly above the black dashed curve in Fig.~\ref{fig_bright_comp}). In the limit of high gain, $\mathcal{F}_\text{C}$ asymptotically approaches $\mathcal{F}_\text{Q}$, but in the low gain limit they differ. 

A simple change of measurement scheme allows us to saturate the QCRB for all gains in both the truncated and conventional SU(1,1) interferometers. We set
\begin{equation}
\hat{M}_{\lambda \text{Q}} = e^{ i \theta_a}\hat{a}_f + e^{ -i \theta_a}\hat{a}^{\dagger}_f + \lambda(e^{ i \theta_b}\hat{b}_f + e^{ -i \theta_b}\hat{b}^{\dagger}_f),
\end{equation}
where $\lambda$ represents a classical gain factor that will multiply mode $b_f$'s homodyne detector output before being added to mode $a_f$. One can evaluate the sensitivity for this measurement scheme, but the expression is lengthy and not given here. If $\lambda \rightarrow \tanh(2r)$, the sensitivity for this scheme saturates the QCRB, which means it is an optimal measurement. For low gains, when $r \to 0$, $\lambda \to 0$, and the measurement is almost entirely due to the detector for mode $a_f$. In this limit, the intensity of mode $b_f$ goes to zero, and the situation approaches that shown in Fig.~\ref{fig_schematics_ph}. Therefore, using $\hat{M}_{\lambda\text{Q}}$ with a bright seed, one can still saturate the QCRB without a need for the second NLO. We plot the sensitivity using $\hat{M}_{\lambda\text{Q}}$ in Figs.~\ref{fig_bright_vs_theta} and~\ref{fig_bright_comp}.

We now consider the effect of losses, starting with those internal to the interferometer ($\eta_{a_i}$, $\eta_{b_i}$ in Fig.~\ref{fig_schematics}). In Fig.~\ref{fig_bright_hd_scaling}, we show how sensitivity in the seeded case changes as a function of gain on a log-log plot. The lowest line shows an ideal, lossless, interferometer that approaches the Heisenberg limit for large gains. Gain is proportional to spontaneous photon number, so $1/\bar{n}^2_s \sim 1/g^2$. We simulate the effect of losses by inserting beamsplitters between NLO 1 and NLO 2, as shown in Fig.~\ref{fig_schematics}. Keeping the losses in each arm the same, we show how not only sensitivity is changed, but how the scaling of sensitivity with gain changes. As losses approach $50$ $\%$, the sensitivity approaches that of the SQL: $1/\bar{n}_s \sim 1/g$ for the range of gain plotted.

\begin{figure}
  \centering
  \includegraphics[width=3.3in]{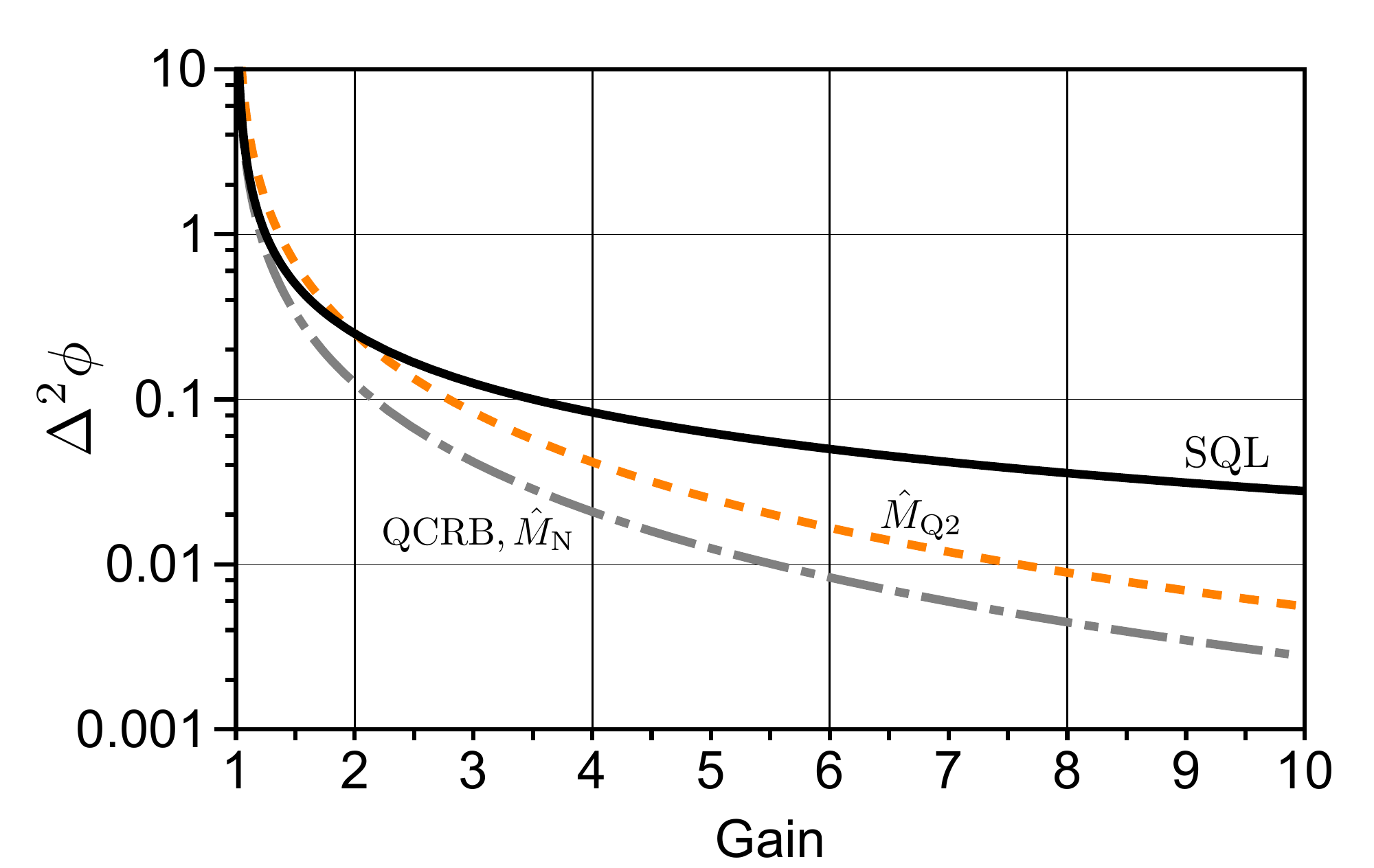}
  \caption
   {Sensitivity as a function of gain for vacuum seeded configurations, optimized over $\phi$ and LO phases. Sensitivity is shown for several experimental configurations: homodyne detection of truncated and conventional SU(1,1) interferometer ($\hat{M}_{\text{Q2}}$) is in orange dashed, intensity detection ($\hat{M}_{\text{N}}$) is in gray dash-dot. The QCRB coincides with the gray dash-dot curve and the SQL is the solid black curve.}
\label{fig_vacuum_comp}
\end{figure}

\begin{figure}
  \centering
  \includegraphics[width=3.3in]{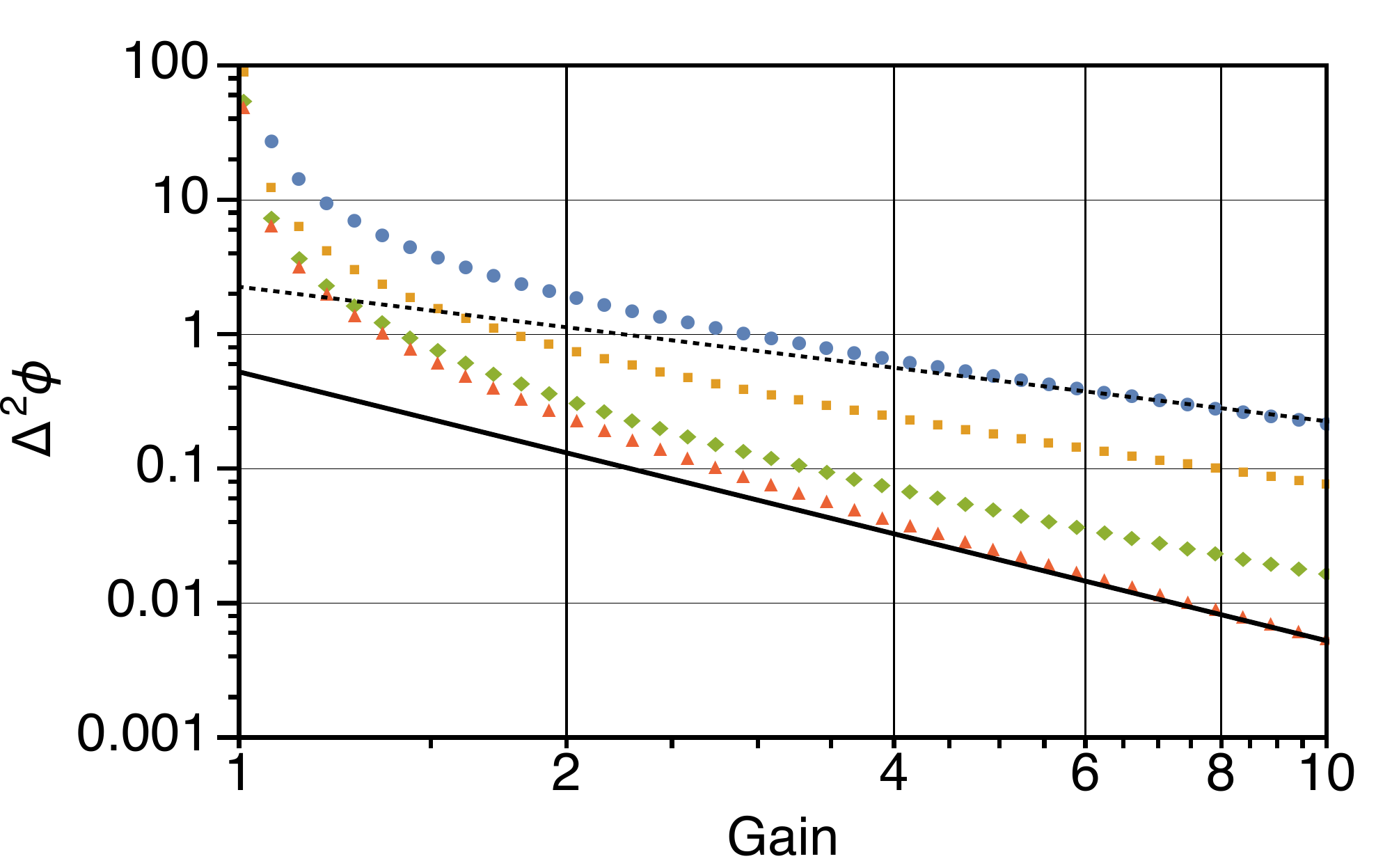}
  \caption
   {Sensitivity as a function of gain for homodyne measurement of vacuum seeded truncated or conventional SU(1,1) interferometer ($\hat{M}_{\text{Q2}}$) and no external loss. Gain is linear in number of spontaneous photons: $\bar{n}_s = 2g - 2$. The blue dots, yellow squares, green diamonds, and orange triangles represent internal transmissions $\eta_{\text{ai}} = \eta_{\text{bi}} = 0.5, 0.75, 0.95$ and $1$, respectively. The black lines are guides for the eye where the dashed has scaling $1/g$ and the solid has scaling $1/g^2$.}
\label{fig_vac_hd_scaling}
\end{figure}

\subsection{Vacuum Seed}
In an SU(1,1) interferometer seeded with vacuum in both modes, the two output modes from the first NLO have a phase relationship even if one beam, by itself, has no phase relationship to any external phase reference. Each mode of the two-mode state, by itself, has thermal statistics. In a quadrature picture, the noise distribution of each mode is centered at the origin and circularly symmetric. Because the expectation value of any quadrature is zero, performing a measurement of the joint quadratures of the two output modes ($\hat{M}_{\text{Q}}$) will not yield a sensitivity. If we calculate the classical Fisher information for this scenario, we see the first term in Eq.~(\ref{eq_CFI}) (the sensitivity) is zero, but the second term is non-zero. In principle, there is more information to extract from the quadrature measurements than what's given by the sensitivity. We can simply define a new measurement that allows access to the information contained in the second term of Eq.~(\ref{eq_CFI}).

To do this, we set 
\begin{equation}
\hat{M}_{\text{Q2}} = (e^{ i \theta_a}\hat{a}_f + e^{ -i \theta_a}\hat{a}^{\dagger}_f + e^{ i \theta_b}\hat{b}_f + e^{ -i \theta_b}\hat{b}^{\dagger}_f)^2,
\end{equation}
 which is a measurement of the noise power in a particular joint quadrature. In this case, we find
\begin{multline}
(\Delta^2 \phi)^{\text{U}}_{\text{Q2}} =  \\
 \frac{1}{2} \csc ^2(\phi -\theta_b) [2 \cos (\phi -\theta_b) +\tanh (r)+\coth (r)]^2.
\end{multline}
As
\begin{equation}
\phi -\theta_b \to \pi -\tan ^{-1}\left[\text{csch}(2 r)\right],
\end{equation}
we achieve the optimum value,
\begin{equation}
(\Delta^2 \phi)^{\text{U}}_{\text{Q2}}\Big |_{\text{min}}  = 2~\text{csch}^2(2 r),
\label{eq_vac_hd_sensi}
\end{equation}
for both the truncated and conventional schemes. We show the sensitivity of the vacuum-seeded scheme as a function of $\phi$ in Fig.~\ref{fig_vacuum_vs_theta}. The optimal phase sensitivity from Eq.~(\ref{eq_vac_hd_sensi}) can still be better than the SQL, but is worse than that of the conventional SU(1,1) interferometer with $\hat{M}_{\text{N}}$. Nevertheless, there may be practical reasons to use homodyne detection in vacuum-seeded experiments, for example, to overcome technical issues such as scattered pump light that reaches optical detectors. We don't describe any measurement with a $\lambda$ correction factor (comparable to $\hat{M}_{\lambda\text{Q}}$) in the vacuum case because the mean intensities of the two modes $a_i$ and $b_i$ are the same.

The comparison among vacuum-seeded schemes as a function of gain is shown in Fig.~\ref{fig_vacuum_comp}. Similar to the bright seeded case, the scaling is badly degraded with loss as shown in Fig.~\ref{fig_vac_hd_scaling}. 

As described above, if we had instead performed a measurement $\hat{M}_{\text{Q}}$, we would find that the sensitivity is very poor (in fact, infinite). However, if we calculate $\mathcal{F}_\text{C}$ using $\hat{M}_{\text{Q}}$, the second term of Eq.~(\ref{eq_CFI}) is nonzero and it would be equal to the sensitivity found in Eq.~(\ref{eq_vac_hd_sensi}). Therefore, for vacuum seeds, the CCRB using $\hat{M}_{\text{Q}}$ is identical to calculating the sensitivity given measurement $\hat{M}_{\text{Q2}}$. 

\section{Sensitivity as a function of input seed}
\label{sec_sensi_seed}
\begin{figure}
  \centering
  \includegraphics[width=3.3in]{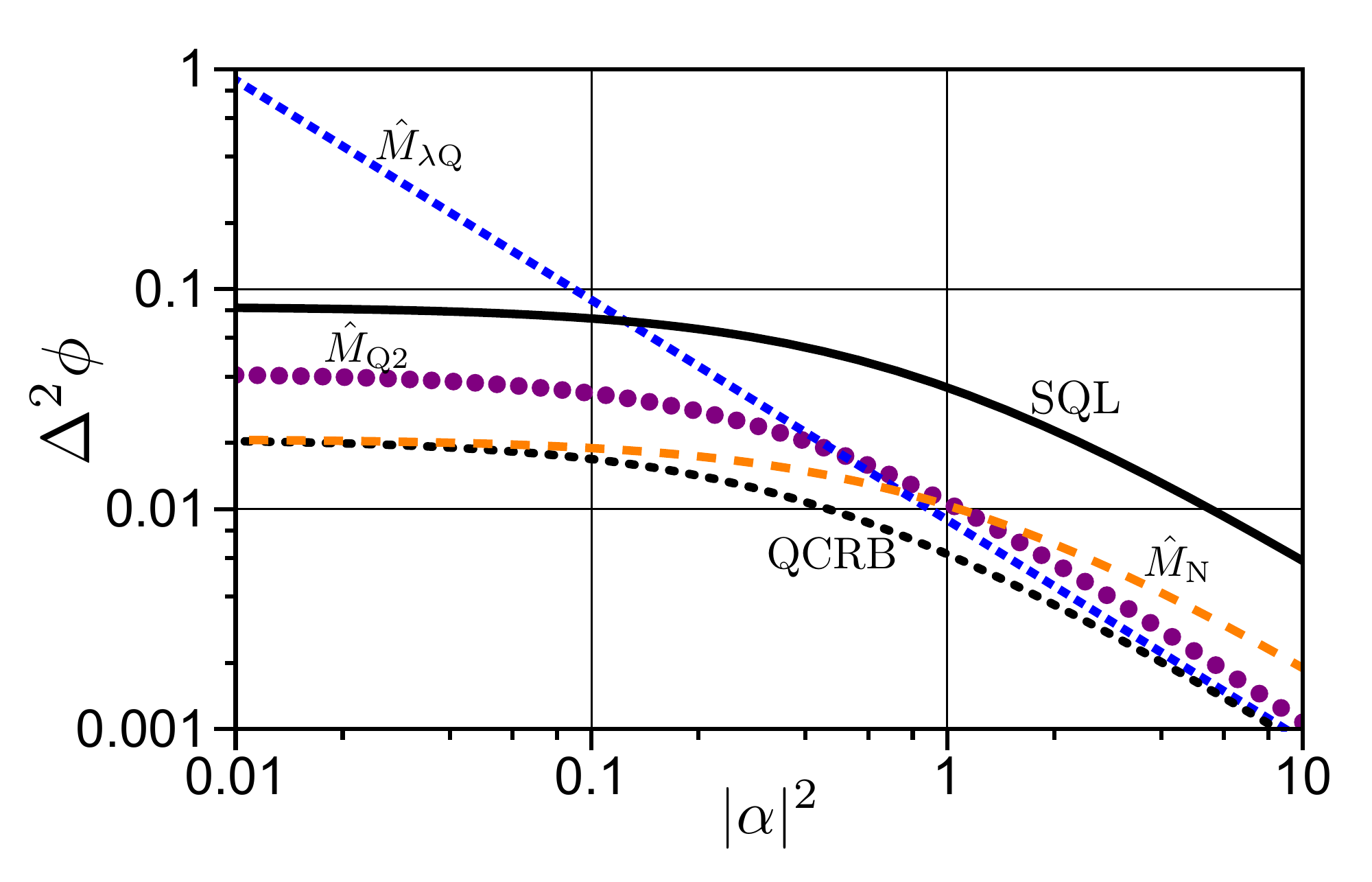}
  \caption
   {Sensitivity as a function of mean input seed photon number with $g=4$, optimized over $\phi$ and LO phases. Sensitivity is shown for several experimental configurations: homodyne detection of truncated SU(1,1) interferometer with classical gain correction ($\hat{M}_{\lambda \text{Q}}$) is dotted blue, homodyne detection of truncated and conventional SU(1,1) interferometer ($\hat{M}_{\text{Q}2}$) is purple circles, and intensity detection of conventional SU(1,1) ($\hat{M}_{\text{N}}$) is dashed orange. The QCRB is the dashed black curve and the SQL is the solid black curve. For small seeds, intensity detection is optimal, and for large seeds homodyne detection is optimal.}
\label{fig_sensi_vs_gamma}
\end{figure}

In the previous sections, we showed that when the input seed is vacuum, the conventional SU(1,1) with intensity measurement $\hat{M}_{\text{N}}$ is optimal. When we have a bright coherent seed, the optimal detection scheme is the truncated or conventional SU(1,1) with homodyne measurement $\hat{M}_{\lambda \text{Q}}$. Clearly, the optimal measurement changes as a function of the input seed brightness. This is shown in Fig.~\ref{fig_sensi_vs_gamma}, where can see that at very low input seed, the intensity measurement ($\hat{M}_{\text{N}}$) approaches the limit set by the quantum Fisher information. At large seeds, the homodyne measurement ($\hat{M}_{\lambda \text{Q}}$) approaches the limit set by the quantum Fisher information. The cross-over point at which photon number measurement and homodyne measurement switch being the optimal measurement is near when $|\alpha|^2 \lesssim 1$.

\section{Compensating for External Loss with Gain}
\label{sec_ext_loss_gain}

\begin{figure}
  \centering
  \includegraphics[width=3.3in]{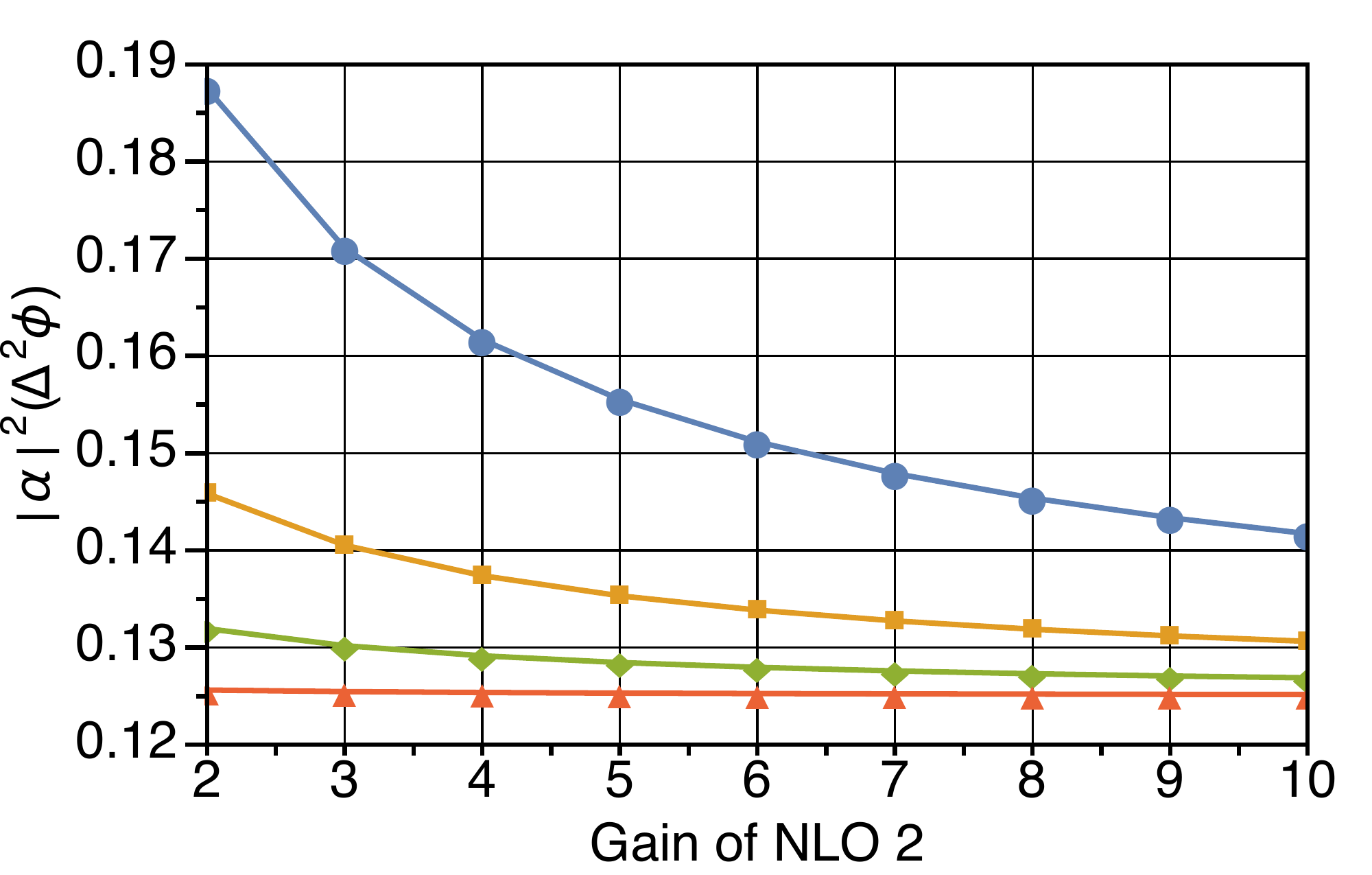}
  \caption{
Sensitivity versus gain of NLO 2 for intensity detection ($\hat{M}_{\text{N}}$) of conventional SU(1,1) interferometer, optimized over $\phi$. NLO 1 has $g=2$ and there is no internal loss. The blue dots, yellow squares, green diamonds, and orange triangles represent external transmissions of $\eta_{\text{ae}} = \eta_{\text{be}} = 0.5, 0.75, 0.9$ and $0.99$, respectively.}
\label{fig_vacuum_number_perfectdet}
\end{figure}

The experimental setups in Figs.~\ref{fig_schematics}a and~\ref{fig_schematics}b assumed that both NLO 1 and NLO 2 had the same gain. Although we claim that the second NLO is not fundamentally needed, the second NLO can be helpful in the case of loss~\cite{Caves1981,Manceau2017}. We distinguish two different types of losses: `external' losses and `internal' losses. Internal losses are shown in Fig.~\ref{fig_schematics} by beamsplitters with transmission $\eta_{\text{ai}}$ and $\eta_{\text{bi}}$, and external losses are shown by $\eta_{\text{ae}}$ and $\eta_{\text{be}}$. The distinction between detection losses and internal losses has been made elsewhere \cite{Marino2012a}. Here we analyze gain imbalance when there are significant external losses. 

The second NLO acts as a phase-sensitive amplifier, which can exhibit noiseless amplification~\cite{Corzo2012,Caves1982}. If we have external losses, it is better to amplify modes $a_i$ and $b_i$ before the loss, assuming the amplifier is noiseless. We show the results in Fig.~\ref{fig_vacuum_number_perfectdet} for the experimental configuration shown in Fig.~\ref{fig_schematics}a, seeded with the vacuum. Increasing the gain of NLO 2 is beneficial if there is significant external loss.

This can be seen analytically if we calculate $\Delta \phi$ for the conventional SU(1,1) interferometer with measurement $\hat{M}_{\text{N}}$ and take $\eta_{\text{ae}}=\eta_{\text{be}}$. We won't reproduce the expression here, but if we take $\lim_{g_2 \rightarrow \infty} \Delta \phi$, where $g_2$ is the gain of NLO 2, the expression is independent of $\eta_{\text{ae}}$. The expression also has the same optimal sensitivity as the no loss case where $\eta_{\text{ae}} = \eta_{\text{be}} = 1$. Therefore, as the second NLO's gain is made arbitrarily large, we overcome any external losses.

\section{Conclusion}

In this work, we have investigated several configurations of SU(1,1) interferometers and compared their phase sensing abilities. We summarize the measurement phase sensitivities compared to the QCRB in Table~\ref{tab_summary}. We have shown that for unseeded SU(1,1) interferometers, optical intensity detection is an optimal scheme that saturates the quantum Cramer-Rao bound. For bright seeded interferometers, we presented a measurement scheme using optical homodyne detection that also saturates the quantum Cramer-Rao bound. This measurement scheme can be implemented using a simplified, ``truncated" version of an SU(1,1) interferometer, which may be more easily implemented in an experiment. In the case of an unseeded interferometer, where the phase sensing beams have no well-defined optical phase, we have analyzed a method for phase measurements with optical homodyne detection. While not achieving the QCRB, this method can still surpass the SQL and may be useful experimentally. We have also shown how one might compensate for external detection losses by increasing the parametric gain of a second nonlinear optical process. These results provide guidelines for the optimal detection schemes that should be used for phase sensing in quantum-enhanced interferometers.

\begin{table}[ht!]
\setlength\extrarowheight{3pt}
\caption{Summary of measurement phase sensitivies compared to the QCRB.}
\begin{tabular}{ |p{0.7cm}||p{3.9cm}|p{3.0cm}| }
 \hline
$\hat{M}$ & $|\alpha|^2\gg 1$ & $|\alpha|^2 = 0$\\ \hline \hline
$\hat{M}_{\text{Q}}$ & saturates QCRB as $g\rightarrow \infty$ & no  phase sensitivity \\  \hline
$\hat{M}_{\text{Q}2}$ &  sub-optimal &  sub-optimal\\ \hline
$\hat{M}_{\lambda\text{Q}}$ & saturates QCRB &  no phase sensitivity\\ \hline
$\hat{M}_{\text{N}}$ &   sub-optimal& saturates QCRB\\ \hline
$\hat{M}_{\text{Nb}}$ &   sub-optimal& saturates QCRB\\ \hline
\end{tabular}
\label{tab_summary}
\end{table}

\begin{acknowledgments}
We gratefully acknowledge discussions with T. Horrom, P. Barberis-Blostein, C. Hermann-Avigliano, E. Tiesinga and E. Goldschmidt. We thank the National Science Foundation and the Air Force Office of Scientific Research.
\end{acknowledgments}


\end{document}